\documentclass[
superscriptaddress,
 amsmath,amssymb,
 aps, twocolumn,
]{revtex4-2}

\usepackage{graphicx}
\usepackage{dcolumn}
\usepackage{bm}
\usepackage{xcolor} 

\usepackage{xcolor,hyperref}
\hypersetup{
   colorlinks,
   linkcolor={blue!50!black},
   citecolor={blue!50!black},
   urlcolor={blue!80!black}
}

\DeclareMathAlphabet{\mathitbf}{OML}{cmm}{b}{it}
\newcommand{\zerovector}{\mathBold 0}
\newcommand{\ket}[1]{|#1\rangle}
\newcommand{\bra}[1]{\langle #1|}
\newcommand{\braket}[2]{\langle #1|#2\rangle}

\newcommand{\calBold}[1]{\mbox{\boldmath${\cal #1}$}}
\newcommand{\mathBold}[1]{\mbox{\boldmath$#1$}}
\newcommand{\dbar}{{\,\,\,\mathchar'26\mkern-12mu d}}

\usepackage{enumitem} 
\usepackage{appendix}

\let\centeredsection\section

\usepackage{etoolbox}
\patchcmd{\section}{\centering}{}{}{}

\let\flushleftsection\section
\newcommand{\sectionscenter}{\let\section\centeredsection}
\newcommand{\sectionsleft}{\let\section\flushleftsection}

\let\centeredsubsection\subsection

\usepackage{etoolbox}
\patchcmd{\subsection}{\centering}{}{}{}

\let\flushleftsubsection\subsection
\newcommand{\subsectionscenter}{\let\subsection\centeredsubsection}
\newcommand{\subsectionsleft}{\let\subsection\flushleftsubsection}

\let\centeredsubsubsection\subsubsection

\usepackage{etoolbox}
\patchcmd{\subsubsection}{\centering}{}{}{}

\let\flushleftsubsubsection\subsubsection
\newcommand{\subsubsectionscenter}{\let\subsubsection\centeredsubsubsection}
\newcommand{\subsubsectionsleft}{\let\subsubsection\flushleftsubsubsection}

\begin{document}

\preprint{revtex4.2 preprint}

\author{Julia A. Giannini}
\email{jagianni@syr.edu}
\affiliation{Department of Physics, Syracuse University, Syracuse, New York 13244, USA}
\affiliation{BioInspired Institute, Syracuse University, Syracuse, New York 13244, USA}
\author{David Richard}
\affiliation{Department of Physics, Syracuse University, Syracuse, New York 13244, USA}
\affiliation{Institute for Theoretical Physics, University of Amsterdam, Science Park 904, Amsterdam, Netherlands}
 \author{M. Lisa Manning}
\affiliation{Department of Physics, Syracuse University, Syracuse, New York 13244, USA}
\affiliation{BioInspired Institute, Syracuse University, Syracuse, New York 13244, USA}%
\author{Edan Lerner}
\email{e.lerner@uva.nl}
\affiliation{Institute for Theoretical Physics, University of Amsterdam, Science Park 904, Amsterdam, Netherlands}

\title{Bond-space operator disentangles quasi-localized and phononic modes in structural glasses}


\date{\today}

\begin{abstract}

The origin of several emergent mechanical and dynamical properties of structural glasses is often attributed to populations of localized structural instabilities, coined \emph{quasilocalized modes} (QLMs). Under a restricted set of circumstances, glassy QLMs can be revealed by analyzing computer glasses' vibrational spectra in the harmonic approximation. However, this analysis has limitations due to system-size effects and hybridization processes with low energy phononic excitations (plane waves) that are omnipresent in elastic solids. Here we overcome these limitations by exploring the spectrum of a linear operator defined on the space of particle interactions (bonds) in a disordered material. We find that this bond-force-response operator offers a unique interpretation of QLMs in glasses, and cleanly recovers some of their important statistical and structural features. The analysis presented here reveals the dependence of the number density (per frequency) and spatial extent of QLMs on material preparation protocol (annealing). Finally, we discuss future research directions and possible extensions of this work. 

\end{abstract}

\maketitle



\setcounter{section}{0}%
\setcounter{subsection}{0}%
\setcounter{figure}{0}    

\section{\label{sec:introII}Introduction}
Glasses and other amorphous solids represent a class of materials that is both relatively commonplace and highly complex. Traditional glasses can be engineered to have desired mechanical and optical properties and have been used in commercial and technological applications for decades \cite{cubuk_structure-property_2017,berthier_theoretical_2011,richard_predicting_2020}. Despite their familiarity, the fundamental physics underlying several common features of glasses is not yet well-understood. For example, thermodynamic and mechanical properties of disordered solids such as the dependence of heat capacity on temperature~\cite{Zeller_and_Pohl_prb_1971} and material response to external deformation~\cite{barrat_yielding_jcp_2004,Fielding_prl_2020_ductile_brittle} vary non-trivially from those of crystalline solids.

While amorphous materials respond elastically to small applied strain, they undergo irreversible structural rearrangement for moderate deformation that is difficult to characterize and predict~\cite{richard_predicting_2020}. In early studies of metallic glasses, researchers identified localized regions of stress-induced plastic deformation that are responsible for yielding behaviors such as shear banding and avalanches \cite{argon_plastic_1979}. Falk and Langer analyzed micromechanical features that give rise to localized irreversible rearrangements in simulated solids, and termed such glassy defects ``Shear Transformation Zones" (STZs) \cite{falk_dynamics_1998}. Recently, significant effort has been put into forming structure-dynamics predictions for the failure behavior of disordered solids \cite{cubuk_structure-property_2017, richard_predicting_2020, scalliet_nature_2019, patinet_connecting_2016, morse_differences_2020, manning_vibrational_2011,plastic_modes_prerc,zohar_prerc}. These works attempt to identify structural precursors to plastic deformation, which occurs when glasses become unstable. 

\begin{figure}[!ht]
\includegraphics[width = 1.\columnwidth]{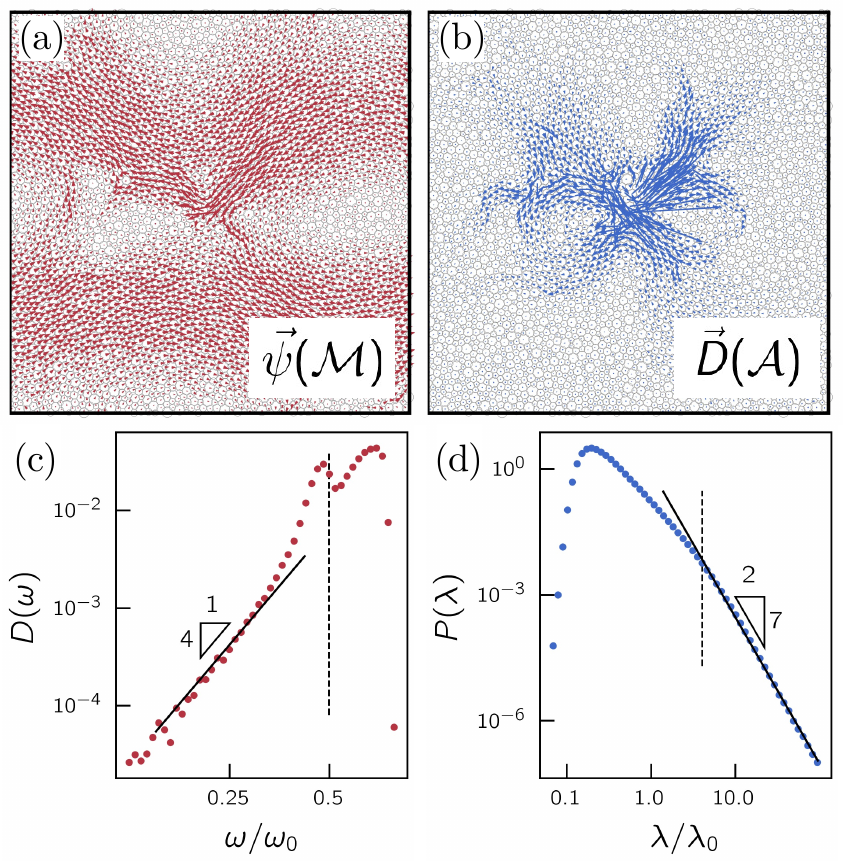}
\caption{A single soft mode in a two-dimensional (2D) computer glass with $N\!=\!4096$ as revealed by \textbf{(a)} the Hessian matrix ${\cal M}$, and by \textbf{(b)} the bond-force-response operator ${\cal A}$, see text for details. The clear phonon hybridization of the Hessian mode of panel (a) is entirely suppressed in the corresponding ${\cal A}$-mode of panel (b), cleanly revealing a \emph{quasilocalized mode}. \textbf{(c)} Harmonic spectrum $D(\omega)$ vs.~frequency $\omega$ for computer glasses in three dimensions (3D), featuring the universal asymptotic $\!\sim\!\omega^4$ scaling. The vertical line marks the expected lowest phonon frequency $\omega_{\text{ph}}\! =\!2 \pi c_s/L$ with $c_s$ denoting the shear wave speed and $L$ denoting the linear size of the glass. All frequencies are divided by $\omega_0 = c_s/a_0$ where $a_0 \sim \rho^{-1/d}$ and $\rho$ is the density of the system (with particle masses set to unity). \textbf{d)} Distribution of eigenvalues $P(\lambda)$ of ${\cal A}$ calculated for the same glasses as (c). All eigenvalues are divided by $\lambda_0 = \omega_0^{-2}$ for sake of comparison with the spectrum of the Hessian. The observed $P(\lambda)\!\sim\!\lambda^{-7/2}$ scaling at large $\lambda$ echos the $\sim\!\omega^4$ scaling of the nonphononic spectrum of (c), as we explain in what follows.}

\label{fig:example_hessian_bond}
\end{figure}

Many methods for detecting STZs in glasses build upon Goldstein's Potential Energy Landscape (PEL) picture \cite{Goldstein1969,Malandro_Lacks,lemaitre2004,plastic_modes_prerc,richard_predicting_2020}; for a disordered solid in $\dbar$ spatial dimensions with a total potential energy $U$ that depends on $N\!\!\dbar$ degrees of freedom, the PEL is a $N\!\!\dbar$-dimensional surface that governs the system's mechanics. It is commonly accepted that the intrinsic structural-mechanical disorder of glasses is manifested in the complexity or ``ruggedness" of the PEL, which features a multiplicity of local minima that is exponential in the number of particles $N$. Mechanically stable configurations sit in local minima of the PEL, and applied deformation, forcing, or thermal activity can push the system into adjacent minima, constituting irreversible particle rearrangements. In the framework of linear response theory, one analyzes local curvatures of the PEL by computing the Hessian matrix $\mathcal{M}\!\equiv\!\frac{\partial ^2 U}{\partial \vec{X} \partial \vec{X}}$, where $\vec{X}$ denotes particle coordinates. Diagonalization of $\mathcal{M}$ gives access to a glass's vibrational modes $\vec{\psi}_l$, with associated vibrational frequencies $\omega_l$, satisfying ${\cal M}\cdot\vec{\psi}_l\!=\!\omega_l^2\vec{\psi}_l$ (setting all masses to unity). Under a restricted set of conditions~\cite{lerner_statistics_2016,bouchbinder_universal_2018,lerner_finite-size_2020}, low-frequency harmonic modes can cleanly localize on small groups of particles and constitute good representations of STZs in glasses~\cite{richard_predicting_2020,richard_simple_2021}. These soft quasilocalized modes (QLMs) are key candidates for two-level systems whose presence can explain the thermodynamic and mechanical anomalies of glassy behavior \cite{Zeller_and_Pohl_prb_1971}.

QLMs are soft excitations that emanate from the structural disorder and mechanical frustration of glasses. A subset of these excitations constitute glasses' carriers of plastic deformation~\cite{richard_simple_2021}. It is now well-established that the density $D(\omega)$ of QLMs --- per frequency and per particle --- follows a universal form, scaling as $\omega^4$, independent of spatial dimension~\cite{kapteijns_universal_2018}, glass preparation protocol~\cite{LB_modes_2019,rainone_pinching_2020}, or microscopic-interaction details~\cite{modes_prl_2020}. However, (linear) continuum elasticity dictates that the vibrational spectrum of a solid must include long-wavelength phononic excitations. Since QLMs and phonons are not necessarily orthogonal, hybridization processes between the two --- as visualized in Fig.~\ref{fig:example_hessian_bond}a --- obscure the important information contained in QLMs~\cite{gartner_nonlinear_2016,bouchbinder_universal_2018}. 

To overcome the aforementioned phonon-hybridization effects on QLMs, novel techniques and computational frameworks have been developed. These include a family of nonlinear excitations~\cite{plastic_modes_prerc,gartner_nonlinear_2016,richard_simple_2021, kapteijns_nonlinear_2020}, termed nonlinear plastic modes (NPMs), that constitute solutions to various nonlinear PEL-based micromechanical equations. Importantly, nonlinear excitations do not hybridize with phonons, and have been shown to converge in terms of structure and energy to harmonic modes in the limit of low frequency and in the absence of phonon hybridization~\cite{plastic_modes_prerc,gartner_nonlinear_2016,kapteijns_nonlinear_2020,lerner_micromechanics_2016}. While nonlinear excitations are reliable representations of QLMs, some of them are challenging to compute, as discussed in Ref.~\onlinecite{richard_simple_2021}. In contrast, pseudoharmonic modes (PHMs) --- introduced and discussed in Ref.~\onlinecite{richard_simple_2021} --- are a type of NPM that rely only on the Hessian matrix and do not require computing high order derivatives of the potential, but are still entirely robust against phonon-hybridizations. However, there is currently no well-established way to obtain the full distribution of QLMs in glasses by computing NPMs. 

It is important to note that material preparation plays an important role in glassy yielding behavior~\cite{lewandowski2001effects,rycroft2012fracture,ozawa2018random,richard2021brittle}. Building on the resemblance between the spatial structure of QLMs in computer glasses and the typical response of a material to applied local force dipoles, recent work has i) identified a characteristic energy scale of QLMs and ii) studied the effect of thermal annealing on the abundance, size, and stiffness of QLMs in computer glasses (Refs.~\onlinecite{lerner_characteristic_2018, rainone_pinching_2020, rainone_statistical_2020}). Understanding these physical properties of QLMs is very important for the efficacy of several theoretical frameworks that make predictions about the elasto-plastic deformation of glasses. For example, STZ theory (Refs.~\onlinecite{falk_dynamics_1998, langer_shear-transformation-zone_2008, langer_shear-transformation-zone_2015}), Soft Glassy Rheology (SGR) (Refs.~\onlinecite{sollich_rheology_1997,sollich_rheological_1998}), and elastoplastic modes (Ref.~\onlinecite{nicolas_deformation_2018}) all rely on the existence of strain-accommodating defects. 

Here, we study the statistics of a bond-force-response operator $\mathcal{A}$ (referred to in what follows as the `bond operator', for brevity) in the context of the properties of soft excitations in structural glasses. In essence, the linear operator $\mathcal{A}$ describes the local strain induced at one point in the material, which results from applying a unit force dipole elsewhere in the material (see precise definition below). By construction, the contribution of phonons to the bond operator is regular, allowing $\mathcal{A}$ to cleanly reveal the populations of QLMs in model glasses. This operator was first introduced in Ref.~\onlinecite{lerner_breakdown_2014}, where it was used to identify the lengthscale associated with dipolar response fields in disordered packings of soft disks. Inspired by Refs.~\onlinecite{lerner_breakdown_2014, rainone_pinching_2020}, we thoroughly study the spectral properties of ${\cal A}$ (see example in Fig.~\ref{fig:example_hessian_bond}), and highlight its utility in revealing the statistical, spatial and energetic properties of soft quasilocalized modes in computer glasses.

This paper is structured as follows: Sec. \ref{sec:glass_models} details the ensemble of model glasses analyzed in our study; Sec. \ref{sec:operator} defines the bond operator and its properties; Sec. \ref{sec:spectrum} presents our scaling predictions and numerical results related to the eigenspectrum of the bond operator, including a discussion of the structural features of bond operator modes and their correspondence to QLMs; Sec. \ref{sec:discussion} includes a discussion of our results in the context of recent work.

\section{\label{sec:glass_models}Computer glass model}

In this work, we employ a model computer glass-former which we refer to as Inverse Power Law (IPL) soft spheres, describing the pairwise interaction between particles. Glassy samples consist of $N$ particles in $\dbar$ spatial dimensions interacting via a radially symmetric pairwise interaction potential given by $u(r_{ij})$, where $r_{ij}$ is the distance between particles $i$ and $j$. Configurations are prepared with periodic boundary conditions and energy-minimized according to the total potential energy $U(\vec{X})=\sum_{\langle i,j\rangle}u(r_{ij})$, where the sum $\langle i,j \rangle$ runs over all (unique) pairs of interacting particles. We note and emphasize that all spectral analyses presented in our work were calculated for three-dimensional (3D) glasses, while some results are obtained for 2D glasses for presentational and visualization purposes. 

The IPL model is a polydisperse soft spheres model in which particles interact via a purely repulsive pairwise potential $u(r_{ij})\!\sim\!r_{ij}^{-10}$. We cut-off and smooth the potential up to 2 derivatives, as described e.g.~in~Ref.~\onlinecite{lerner_mechanical_2019}, where it is also explained how we handled sample-to-sample finite-size effects that emanate from the random drawing of the effective particle sizes from a fat-tailed distribution~\cite{ninarello_models_2017}. The configurations we analyze in this study were first equilibrated at a very broad range of \emph{parent temperatures} $T_{\rm p}$ using the SWAP Monte Carlo algorithm~\cite{berthier_equilibrium_2016, ninarello_models_2017,gutierrez_static_2015} that allows for extreme supercooling. Glassy configurations were formed by a conjugate gradient minimization~\cite{mackay_macopt_2004} of equilibrium configurations. Our glassy ensembles for each parent temperature $T_{\rm p}$ consist of 10,000 independent configurations of $N\!=\!2000$ particles in $\!\!\dbar\!=\!3$ dimensions. In what follows, all dimensionful observables are reported in simulational-units as spelled out in Ref.~\onlinecite{lerner_mechanical_2019}

\section{\label{sec:operator}bond-force-response operator}

\subsection{Definitions and formalism}
The bond operator described in detail below is defined in and acts upon the space of interacting pairs or \emph{bonds}. We restrict the discussion to systems for which this \emph{bond-space} is a vector space of dimension $N_{\rm b}\!>\!N\!\!\dbar$, i.e.~larger than the system's configuration space, of dimension $N\!\!\dbar$. Each component in bond space pertains to a single \emph{pair} of interacting particles. Below we follow the notation convention that, unless otherwise specified, lowercase variables (such as the pairwise distances $r_{\alpha}$ between interacting particles) represent bond-space quantities, while uppercase variables (such as particle coordinates $\vec{X}$) represent coordinate-space quantities. Vectors or operators in bond-space are indexed with Greek letters, while those in coordinate-space are indexed with Latin letters~\cite{lerner_breakdown_2014,lerner_simulations_2013}. We will use vector and bra-ket notation interchangeably as it is convenient. 

We consider first the change $\delta r_{\alpha}$ in the length of the $\alpha^{\mbox{\footnotesize th}}$ bond (consisting of particles $i$ and $j$), which results from imposing a (small) displacement $\delta \vec{R}$ to particle coordinates. To first order in the imposed displacement's magnitude, the distance $r_\alpha$ is extended or compressed by
\begin{equation}\label{eq:S_transformation_example}
    \delta r_{\alpha} \simeq \hat{X}_{ij} \cdot \left( \delta \vec{R}_j - \delta \vec{R}_i\right)
\end{equation}
where $\vec{X}_{ij}\!\equiv\!\vec{X}_j\!-\!\vec{X}_i$ is the difference vector that extends from particle $j$ to particle $i$, and $\hat{X}_{ij}\!\equiv\!\vec{X}_{ij}/|\vec{X}_{ij}|$ is the corresponding unit vector. Eq.~(\ref{eq:S_transformation_example}) constitutes a linear transformation from coordinate-space vectors to bond-space vectors via the $N_b\!\times\!N\!\!\dbar$ dimensional operator $\mathcal{S}$, defined precisely as 
\begin{equation}
    \mathcal{S}_{\alpha,k} = \frac{\partial r_\alpha}{\partial \vec{X}_k}\,.
    \label{eq:S_definition}
\end{equation}
Using this definition, it is convenient to express Eq.~(\ref{eq:S_transformation_example}) for the vector $\vert \delta r \rangle $ containing the extension/compression of all bonds in the system in bra-ket notation as~\cite{lerner_unified_2012}
\begin{equation}
    \vert \delta r \rangle = \mathcal{S} \vert \delta R \rangle.
\end{equation}
Furthermore, to map a bond-space vector to coordinate space, we apply the transpose of $\mathcal{S}$, $\mathcal{S}^T$. For example, if we have a set of interparticle forces $\vert f \rangle$ and wish to obtain the net force on each particle $\vert F \rangle$, we simply compute $\vert F \rangle = \mathcal{S}^T \vert f \rangle$. Refs.~\onlinecite{lerner_breakdown_2014, lerner_simulations_2013} discuss the utility of $\mathcal{S}$ and $\mathcal{S}^T$ further. We note that these operators are referred to elsewhere as the compatibility and equilibrium or rigidity matrices respectively \cite{lubensky_phonons_2015}.

\begin{figure}
\includegraphics[width = 1.0\columnwidth]{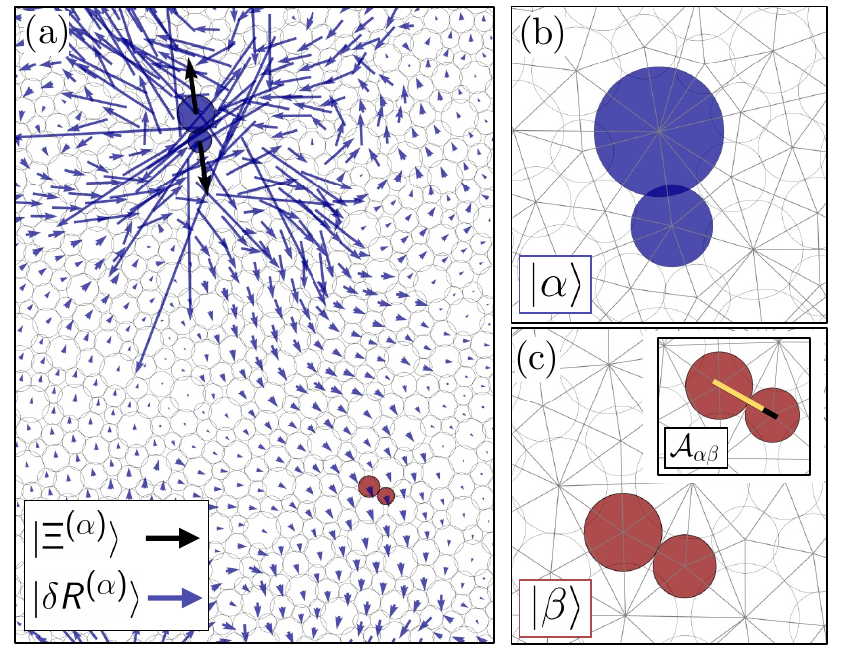}
\caption{Bond operator formalism. For an example 2D IPL system: \textbf{a)} Coordinate space unit dipole on $\alpha$, $\vert \Xi^{(\alpha)} \rangle$ and displacement response $\vert \delta R^{(\alpha)} \rangle$ \textbf{b)} Contact network around $\alpha$ \textbf{c)} Contact network around $\beta$. Inset: $\mathcal{A}_{\alpha \beta}$ measures the change in length of bond $\beta$ in response to an applied dipolar force on $\alpha$.}
\label{fig:formalism}
\end{figure}

We consider next the change in length of bond $\beta$ that results from applying a unit dipolar force to the $\alpha^{\mbox{\footnotesize th}}$ bond~\cite{lerner_breakdown_2014}. A schematic of the relevant objects to this computation in an example 2D system are shown in Fig.~\ref{fig:formalism}. We define the bond-space vector $\vert \alpha \rangle$ to contain all zeros, except for the entry corresponding to the $\alpha^{\mbox{\footnotesize th}}$ bond whose value is unity. A particular bond $\alpha$ is highlighted in panels (a) and (b) of Fig.~\ref{fig:formalism}. Operating with ${\cal S}^T$ on $\ket{\alpha}$ corresponds to a coordinate-space unit dipole $\ket{\Xi^{(\alpha)}}$ such as that depicted in Fig \ref{fig:formalism}a, namely,
\begin{equation}
    \ket{\Xi^{(\alpha)}} = {\cal S}^T\ket{\alpha}\,.
    \label{eqn:coord_ket_1}
\end{equation}
The (linear) displacement response $\vert \delta R^{(\alpha)}\rangle$ to such an applied force is simply
\begin{equation}
    \vert \delta R^{(\alpha)} \rangle = \mathcal{M}^{-1}\ket{\Xi^{(\alpha)}} = {\cal M}^{-1}\mathcal{S}^T \vert \alpha \rangle,
\end{equation}
as shown for example in Fig.~\ref{fig:formalism}a. Here ${\cal M}^{-1}$ should be understood as the pseudo-inverse of ${\cal M}$. Last, we project the displacement response field $\ket{\delta R^{(\alpha)}}$ onto a dipole constructed on the $\beta^{\mbox{\footnotesize th}}$ bond (Fig \ref{fig:formalism}c), to obtain the $(\alpha,\beta)$ element of the $N_{\rm b}\!\times\!N_{\rm b}$-dimensional, symmetric and positive semi-definite bond-force-response operator ${\cal A}$. This element is thus given by 
\begin{equation}
{\cal A}_{\alpha\beta} \equiv \braket{\Xi^{(\beta)}}{\delta R^{(\alpha)}} = \bra{\alpha}{\cal S}{\cal M}^{-1}{\cal S}^T\ket{\beta}\,
\label{eqn:operator_element_1}
\end{equation}
which is schematically shown in the inset of Fig.~\ref{fig:formalism}c, where $\beta$ is extended by a small amount due to the deformation induced by the applied force on $\alpha$.

In the discussion that follows, it will be useful to consider an expression for the bond operator in terms of the full eigenmode decomposition of $\mathcal{M}$:
\begin{equation}
    \mathcal{A} =  \sum_{l=1}^{N\!\!\dbar - \dbar}\frac{\mathcal{S}\vert \psi_l \rangle \langle \psi_l \vert\mathcal{S}^{T}}{\omega^2_l} \, ,
    \label{eqn:A_hess_sum}
\end{equation}
where $\omega_l$ and $\vert \psi_l \rangle$ are the eigenvalues and corresponding eigenvectors of the Hessian. Particularly, writing $\mathcal{A}$ in this way illustrates how its spectral properties can be understood from that of $\mathcal{M}$. We proceed by noting a few additional features of $\mathcal{A}$ that are relevant to its utility in connecting micromechanical information to overall material behavior.

\subsection{Zero-modes of ${\cal A}$ are states of self-stress }

Bond-space vectors $\vert \eta \rangle$ that belong to the left-nullspace of $\mathcal{S}$ represent states of self stress (SSS): sets of bond extensions/compressions that do not alter the state of force balance of a system, but still introduce mechanical stresses \cite{lerner_quasilocalized_2018}. That is, SSS satisfy $\mathcal{S}^{T}\vert \eta \rangle = 0$. Given the definition of $\mathcal{A}$ above (Eqs.~\ref{eqn:operator_element_1} and \ref{eqn:A_hess_sum}), its zero modes will then be exactly the SSS of the system. For granular packings near the unjamming transition, Maxwell constraint counting implies that $\text{\# SSS} \simeq N \delta z/2$ where $\delta z$ is the number of excess contacts (past isostaticity) per particle~\cite{lubensky_phonons_2015}.

\subsection{${\cal A}$ within continuum linear elasticity}
It is possible to write an exact expression for ${\cal A}$ within linear continuum elasticity, however it is more insightful and less cumbersome to spell out scaling arguments to highlight its expected long-wavelength properties. To this aim, we consider the elastic Green's function $\calBold{G}(\vec{r})\!\sim\!r^{-(\dbar-2)}$ (in $\dbar\!>\!2$ dimensions) of a linear-elastic, homogeneous, and isotropic solid~\cite{landau_lifshitz_elasticity}. The displacement field due to a dipole of length $a_0$ scales as $a_0\nabla G\!\sim\!a_0r^{-(\!\!\dbar-1)}$. The strain field at distances $r$ from the imposed dipole is thus expected to scale as $a_0^2\nabla^2 G\!\sim\!a_0^2 r^{-\!\!\dbar}$. Noticing that $\mathcal{A}$ behaves as the gradient of the displacement field that results from applied force dipoles, we conclude that 
\begin{equation}\label{eq:A_scaling_continuum}
    |{\cal A}|(\vec{r})  \sim r^{-\!\!\dbar}\:.
\end{equation}
The scaling relation given by Eq.~(\ref{eq:A_scaling_continuum}) was validated numerically in Refs.~\cite{lerner_breakdown_2014,plastic_modes_prerc}. The $r^{-\!\!\dbar}$ decay of interactions is indeed expected for interacting dipoles, and forms the basis of lattice models of glassy excitations~\cite{Gurevich2003,Gurevich2007,rainone_mean-field_2021}, lending further support to the relevance of ${\cal A}$ in the present discussion.

\section{\label{sec:spectrum}Spectral properties of the bond operator}

\subsection{Low-frequency contributions to the bond operator}
As stated in the introduction, our goal behind studying the bond operator ${\cal A}$ is to propose a systematic route to overcome the hybridization of phonons with QLMs, in order to cleanly access the latter. It is convenient to assess the contributions of low-frequency modes of $\mathcal{M}$ to ${\cal A}$ by writing ${\cal M}^{-1}$ in its eigenbasis as in Eq.~\ref{eqn:A_hess_sum}. In this subsection, we examine the relative contributions of QLMs and phonons to the construction of the bond operator. 

\subsubsection{Phonon contributions to ${\cal A}$}

Here, we show that phononic modes of the Hessian have a regular contribution to $\mathcal{A}$. Viewing the representation of the inverse Hessian in  Eq.~\ref{eqn:A_hess_sum} and recalling Eqs.~\ref{eqn:coord_ket_1} and \ref{eqn:operator_element_1}, we note that computing elements of $\mathcal{A}$ involves projecting the eigenmodes of the Hessian onto pairs of coordinate-space  dipole vectors, $\vert \Xi^{(\alpha)}\rangle$ and $\vert \Xi^{(\beta)}\rangle$. Since phononic modes of $\mathcal{M}$ are extended and vary slowly in space, we expect them to have small projections onto local dipoles. 

More specifically, consider the following expression for the contributions of phonons $\vert \Psi_{\text{ph}, l}  \rangle$ to $\mathcal{A}$:
\begin{equation}
    \mathcal{A}_{\text{ph}} = \sum_{\text{ph}, l} \frac{ \mathcal{S} \vert \Psi_{\text{ph}, l}  \rangle \langle \Psi_{\text{ph}, l}  \vert \mathcal{S}^T}{\omega_{\text{ph}, l}^2}
    \label{eq:phonon_cont}
\end{equation}
where $\omega_{\text{ph}, l}$ is the eigenfrequency associated with $\vert \Psi_{\text{ph}, l}  \rangle$. According to Eqs.~\ref{eq:S_transformation_example} and \ref{eq:S_definition}, $\mathcal{S} \vert \Psi_{\text{ph}, l}  \rangle$ pickes up \emph{local} differences of the wave-like mode $\vert \Psi_{\text{ph}, l}  \rangle$, and is thus expected to scale as the spatial gradient of $\vert \Psi_{\text{ph}, l}  \rangle$, namely $\mathcal{S} \vert \Psi_{\text{ph}, l}  \rangle\! \sim\!a_0 \nabla \vert \Psi_{\text{ph}, l}  \rangle$. Polarization vectors in phonons vary on the scale of their wavelength, so we conclude that $|\mathcal{S} \vert \Psi_{\text{ph}, l}  \rangle|\! \sim \!\omega_{\text{ph}, l}$. Thus, the numerator of Eq.~\ref{eq:phonon_cont} (representing the projection of $\mathcal{M}$ eigenmodes onto dipoles) is on the order of $\omega_{\text{ph}, l}^2$ and $\mathcal{A}_{\text{ph}} \sim \mathcal{O}(1)$. We thus conclude that phononic contributions to $\mathcal{A}$ are regular.

\subsubsection{QLM contributions to ${\cal A}$ }

Similarly to the above discussion, we now consider the contributions of low-frequency QLMs to $\mathcal{A}$. As observed in Refs.~\onlinecite{rainone_pinching_2020,lerner_characteristic_2018,rainone_statistical_2020}, the structure of QLMs is characterized by a disordered core and algebraically decaying field. In contrast to plane wave excitations, the disordered core of a QLM features highly nonaffine displacements, which, in turn, are expected to have large projections on local dipoles $\vert \Xi^{(\alpha)}\rangle$. This implies that contributions to $\mathcal{A}$ from QLMs are dominant compared to other excitations. Consider the sum

\begin{equation}
    \mathcal{A}_{\text{QLM}} = \sum_{\text{QLM}, l} \frac{ \mathcal{S} \vert \Psi_{\text{QLM}, l}  \rangle \langle \Psi_{\text{QLM}, l}  \vert \mathcal{S}^T}{\omega_{\text{QLM}, l}^2}
\end{equation}
where $\vert \Psi_{\text{QLM}, l}  \rangle$ is a QLM with frequency $\omega_{\text{QLM}, l}$. Given the propensity of QLMs for extending/compressing pairwise bonds in their cores, we conclude that $\mathcal{S} \vert \Psi_{\text{QLM},l} \rangle \sim \mathcal{O}(1)$. In contrast to phononic contributions to $\mathcal{A}$, we expect that QLMs of frequency $\omega_{\text{QLM}}$ contribute terms of order $\omega^{-2}_{\text{QLM}}$ to $\mathcal{A}$. 

Importantly, the considerations described above lead us to expect $\mathcal{A}$ to predominantly contain information about disorder-related soft modes in model glasses that is un-obscured by phononic modes. Thus, we expect that the eigen-spectrum of the bond operator should cleanly reflect the full distribution of QLMs in the system. 

\subsection{Eigenvalue distribution of ${\cal A}$}

\subsubsection{High-$\lambda$ scaling of bond operator spectrum}
As previously noted, in the framework of using the harmonic approximation to identify glassy instabilities, one is generally interested in the low-frequency regime of the eigenspectrum of the Hessian, where universal $D(\omega) \sim \omega^4$ scaling is prevalent \cite{kapteijns_universal_2018}. Since $\mathcal{A}$ is directly related to the inverse Hessian, in the results that follow we will equivalently be interested in studying the high-eigenvalue scaling of $\mathcal{A}$ (Fig.~\ref{fig:example_hessian_bond}d). Examining Eq.~\ref{eqn:A_hess_sum}, we see that the eigenvalues $\lambda$ of $\mathcal{A}$ should scale as the inverse squared frequencies of the Hessian: $\lambda\! \sim \!\omega^{-2}$. Applying a transformation of variables to convert from $D(\omega)\sim \omega^4$ to some $P(\lambda)$, we obtain:
\begin{equation}
    P(\lambda) \sim D\big(\omega(\lambda)\big)|d\omega/d\lambda| \sim \lambda^{-\frac{7}{2}}.
    \label{eqn:p_lambda_predict}
\end{equation}
More generally, for $D(\omega)\!\sim\!\omega^\gamma$, we equivalently have $P(\lambda)\!\sim\! \lambda^{-\frac{\gamma + 3}{2}}$. Thus, we have formed a simple scaling prediction for the bond operator eigenspectrum. As we will see, this scaling is robust in the spectrum of $\mathcal{A}$ for a variety of computer glass ensembles. Last, it is important to note that the transformation of variables performed here preserves the prefactor $A_{\rm g}$ of the $\sim\!\omega^4$ scaling (i.e.~$D(\omega)\!=\!A_{\rm g}\omega^4$), and so we expect to observe $P(\lambda)\!\sim\!A_{\rm g}\lambda^{-7/2}$, as discussed extensively below.

\subsubsection{Results: bond operator spectrum}

\begin{figure}[!ht]
\includegraphics[width = 1.0\columnwidth]{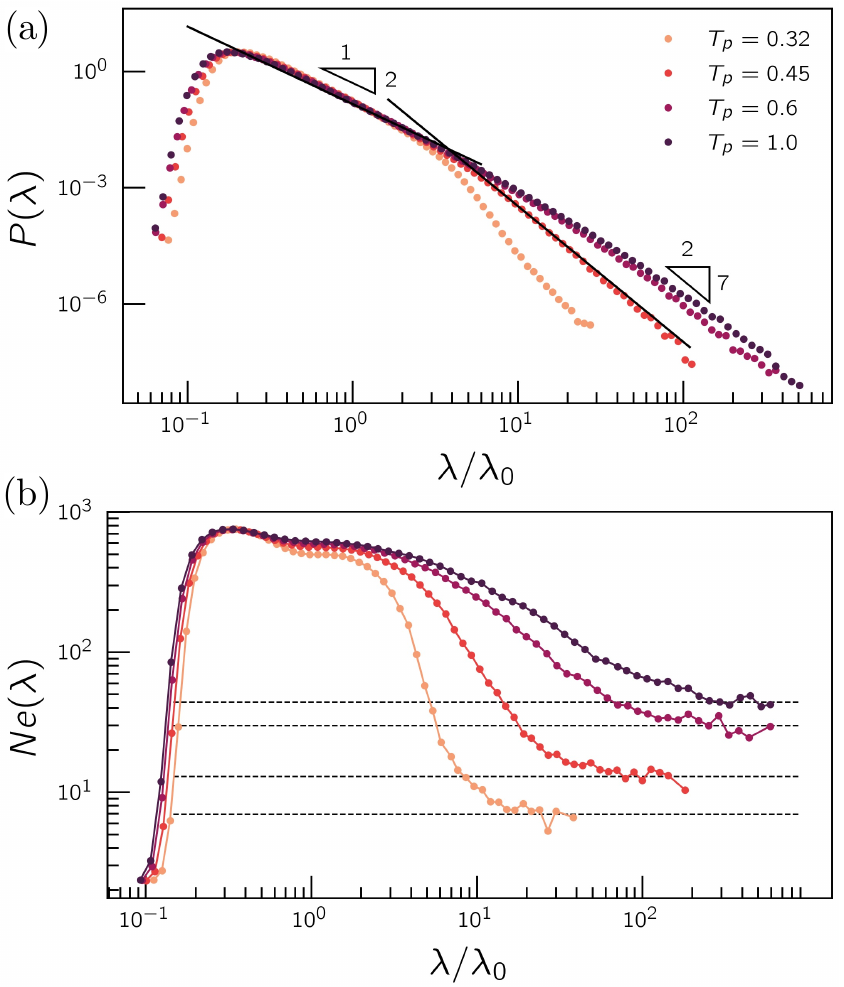}
\caption{Spectrum of bond operator for varying material preparation protocols. \textbf{a)} Distribution of $\mathcal{A}$ eigenvalues for four different $T_{\rm p}$. Darker colors (top) represent deeper annealing. The scale triangles depict the predicted high-$\lambda$ $P(\lambda) \sim \lambda^{-7/2}$ power law and mid-$\lambda$ $P(\lambda) 
\sim \lambda^{-2}$ scaling respectively. \textbf{b)} Participation ratio of particle force modes multiplied by $N$ as a function of $\lambda$ for varying $T_{\rm p}$, color scale as in (a). We use the plateau values of $Ne$ in the high-$\lambda$ regime to extract QLM length scales $\xi$ from $\mathcal{A}$ modes.}
\label{fig:operator_spectrum_1}
\end{figure}

To study the full distribution of disorder-related modes in our model glasses, we computed and diagonalized $\mathcal{A}$ for 10,000-configuration ensembles as noted above, prepared in $\!\!\dbar\!=\!3$ with $N\!=\!2000$ particles at four different parent temperatures $T_p\! \in\! \left\{0.32, 0.45, 0.60, 1.00 \right\}$. Fig.~\ref{fig:operator_spectrum_1}a shows the distributions of eigenvalues for these samples. The high-$\lambda$ regime robustly displays the $\lambda^{-7/2}$ scaling predicted in Eq.~\ref{eqn:p_lambda_predict}, especially for less-deeply annealed glasses such as those with $T_{\rm p}\! \gtrsim\! 0.5$. We also observe that the prefactor $A_{\rm g}$ associated with the $\lambda^{-7/2}$ power law increases with larger $T_{\rm p}$. This is expected, as previous work has shown that the overall number of QLMs depends strongly on equilibrium parent temperature \cite{ lerner_characteristic_2018,LB_modes_2019,rainone_pinching_2020}. We will address this point thoroughly to follow.

The clarity of the $P(\lambda)\!\sim\! \lambda^{-7/2}$ scaling observed here contrasts typical $D(\omega)\! \sim\! \omega^4$ spectra, which are often obscured by the presence of phononic excitations (as also seen in Fig.~\ref{fig:example_hessian_bond}c). Phonons in the spectrum of $\mathcal{M}$ in finite-size solids introduce a system-size dependence that does not exist in that of $\mathcal{A}$; for example, the typical frequency of the first phonons in a solid of size $L$ is $\omega_{\text{ph}}\! =\! 2 \pi c_s/L$ where $c_s$ is the shear wave speed. In finite-size solids the lowest-frequency phonons usually appear in quantized bands~\cite{bouchbinder_universal_2018}, rendering $D(\omega)$ explicitly system-size dependent. This is not the case for the spectrum of $\mathcal{A}$, which is system-size independent as demonstrated in Appendix~\ref{sec:append_system_size}. We emphasize that the scaling behaviors of $D(\omega)$ and $P(\lambda)$ are equivalent, but the bond operator recovers the result more cleanly than the Hessian alone.

We finally note that, in the mid-$\lambda$ regime of the bond operator spectrum, we observe clear $P(\lambda)\! \sim\! \lambda^{-2}$ scaling independent of $T_{\rm p}$, which is reminiscent of $D(\omega)\! \sim\! \omega$ scaling observed at intermediate frequencies in the density of states of lattice models for interacting glassy QLMs~\cite{Gurevich2003,Gurevich2007,rainone_mean-field_2021}. This observation further elucidates $\mathcal{A}$'s utility for identifying and analyzing non-phononic excitations.

\begin{figure*}[ht!]
\includegraphics[width = 1.0\linewidth]{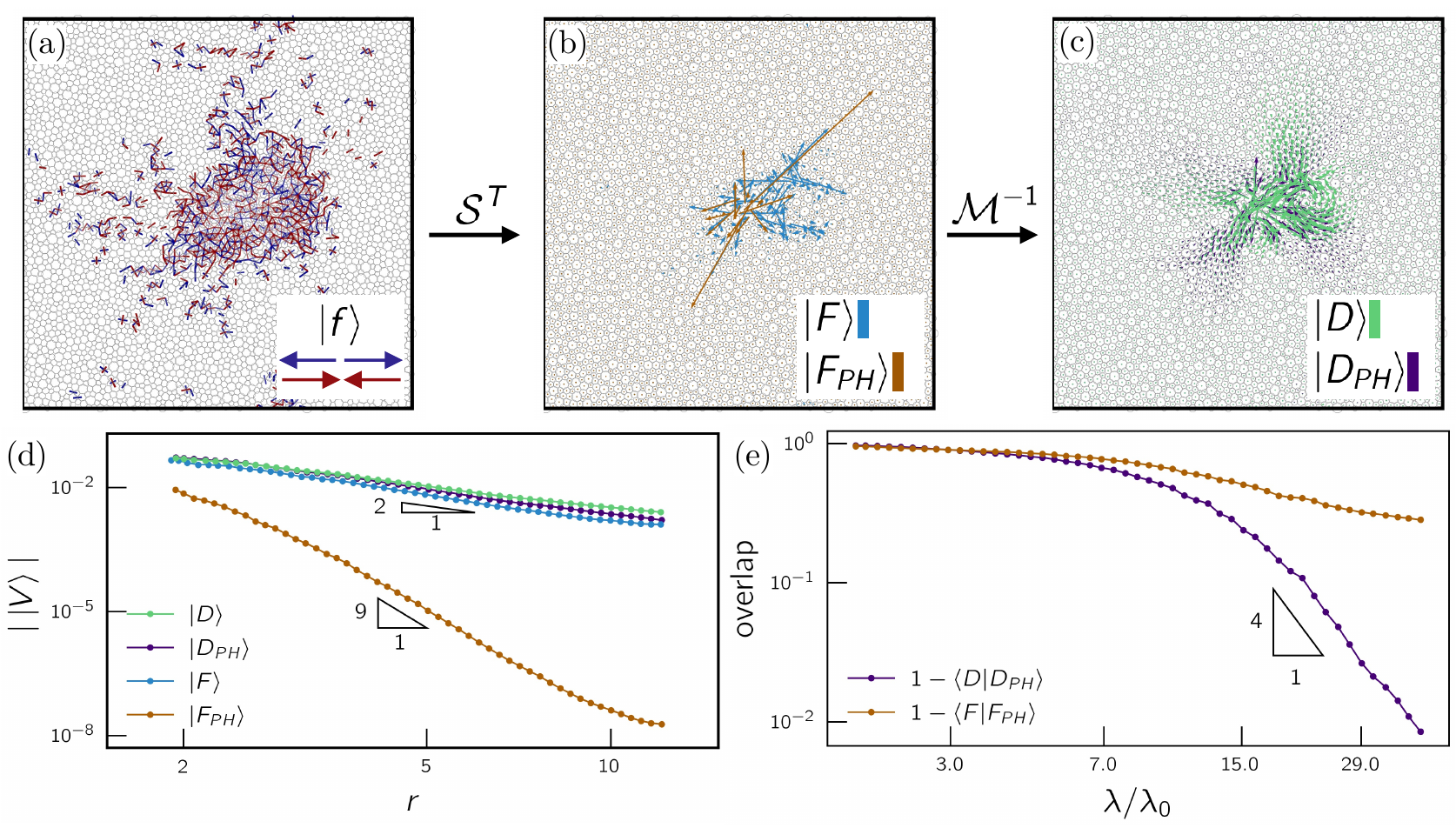}
\caption{Eigenmodes of $\mathcal{A}$ and their correspondence to PHMs. \textbf{a)} high-$\lambda$ bond force mode $\vert f \rangle$ obtained by diagonalizng $\mathcal{A}$. Red-colored bonds represent extension and blue-colored bonds represent compression. Line opacity and thickness are adjusted to reflect the magnitude of extension or compression, and we color only the top $10\%$ (in magnitude) of bonds. \textbf{b)} Particle force mode $\vert F \rangle$ (blue) corresponding to the bond force mode in (a) and related psuedo-harmonic force mode $\vert F_{PH} \rangle$ (orange). \textbf{c)} Particle displacement mode $\vert D \rangle$ (green) corresponding to the particle force mode in (b) and related psuedo-harmonic displacement mode $\vert D_{PH} \rangle$ (purple). \textbf{d)} Spatial decay profiles of polarization vector magnitudes as a function of distance from disordered core. Colored points for each type of mode are the same as in (a)-(c). \textbf{e)} Convergence of particle displacement and particle force modes to their corresponding PHMs. The results shown in (d) and (e) are for our ensemble of 3D IPL glasses, whereas (a)-(c) are presented in 2D for ease of visualization.}
\label{fig:2D_modes_decay_2}
\end{figure*}

\subsubsection{Results: localization of high-$\lambda$ eigenmodes}

An important feature to observe in disordered solids is the presence of a length scale associated with QLM's core size~\cite{rainone_pinching_2020}. Toward this goal, we analyze the participation ratio profile of bond operator eigenmodes as a function of their associated eigenvalues $\lambda$. Specifically, we use coordinate-space modes derived from high-$\lambda$ modes of the bond operator (which we will define as ``particle force modes" and discuss in detail below) to compute participation via
\begin{equation}
    e(\vec{F}) \equiv \frac{\left( \sum_i \vec{F}_i \cdot \vec{F}_i \right)^2 }{N \sum_i \left(\vec{F}_i \cdot \vec{F}_i \right)^2},
\end{equation}
where the sums are over particle indices and $\vec{F}_i$ is the $\!\!\dbar$ dimensional vector prescribing the force on $i$ in the aforementioned coordinate-space mode. This analysis is summarized in Fig.~\ref{fig:operator_spectrum_1}b, which shows $Ne(\lambda)$ for the same group of glass ensembles as in Fig.~\ref{fig:operator_spectrum_1}a. The low-$\lambda$ regime is characterized by very localized modes that quickly increase in participation with growing $\lambda$. These correspond to high-stiffness excitations such as Eschlby-like deformations on a small number of particles, transitioning to more extended modes that apply small forces to many particles. See Appendix \ref{sec:append_modes} for further details regarding the participation of intermediate-$\lambda$ $\mathcal{A}$ eigenmodes. 

More interestingly, for high $\lambda$, the coordinate-space $\mathcal{A}$ modes become localized and strongly resemble QLMs, reaching a plateau in participation as indicated by the horizontal lines in Fig.~\ref{fig:operator_spectrum_1}b. By identifying the participation for which the $Ne$ profile reaches $15\%$ above its plateau value, we extract the typical number $n$ of particles that participate in QLM-like excitations obtained from $\mathcal{A}$, usually $\sim 10$s of particles. Each $n$ thus gives a characteristic QLM length scale $\xi$ via 
\begin{equation}
    \xi = a_0n^{\frac{1}{\!\!\dbar}}
    \label{eqn:qlm_lengthscale}
\end{equation}
where $a_0\!\equiv\!(V/N)^{1/\!\!\dbar}$ is a characteristic interparticle distance. As we will discuss, the length-scale associated with high-$\lambda$ eigenmodes decreases significantly with deeper annealing, in agreement with other methods for investigating the interplay of material preparation and QLM properties \cite{rainone_pinching_2020,rainone_statistical_2020,lerner_characteristic_2018}.

\subsection{Eigenvectors of ${\cal A}$ and their connection to QLMs}

\subsubsection{Relevant modes and spatial decay}

In the discussion that follows, we will define a family of modes that can be derived from the bond operator and briefly formulate scaling arguments that predict their characteristic spatial decay. \vspace{2mm}

\paragraph{Bond force modes.} By diagonalizing $\mathcal{A}$ we obtain its eigenvectors, which we will refer to as \textit{bond force modes}, denote $\vert f \rangle$, and interpret as fields of bond extension/compression forces. Bond force modes are $N_{\rm b}$-dimensional, and can also be understood as fields that describe local strain on the scale of particle bonds. 

\paragraph{Particle force modes.} To instead study bond force modes in coordinate space, we compute \textit{particle force modes} $\vert F \rangle$, via $\vert F \rangle = \mathcal{S}^T \vert f \rangle$. Particle force modes are $Nd$-dimensional and can be interpreted as applied forces on each particle that constitute an equivalent deformation to the corresponding bond force mode. 

\paragraph{Particle displacement modes.} Recall that harmonic vibrational modes (of the Hessian) and QLMs are typically viewed as putative displacement fields about a mechanically stable configuration that can represent structural instabilities or loci of potential plastic yielding. Thus, we compute $N\!\!\dbar$-dimensional \textit{particle displacement modes} $\vert D \rangle$ by applying the inverse Hessian to a particle force mode: $\vert D \rangle = \mathcal{M}^{-1} \mathcal{S}^{T} \vert f \rangle$. 

\paragraph{Pseudoharmonic displacement modes.} As previously noted, several recent works (Refs.~\onlinecite{lerner_micromechanics_2016, plastic_modes_prerc, gartner_nonlinear_2016, kapteijns_nonlinear_2020, richard_predicting_2020, richard_simple_2021}) have formulated and studied NPMs and PHMs as faithful representations of structural instabilities in model glasses. Significantly, these modes resist hybridization with phononic excitations, which is a notable shortcoming of traditional harmonic vibrational modes \cite{richard_simple_2021}. Here, we focus on PHMs as a benchmark for using modes of the bond operator to identify QLMs in model glasses. We first provide background information about formulating PHMs, and then define the corresponding PHMs to particle displacement modes and particle force modes. 

Pseudo-harmonic modes (PHMs) $\vec{\pi}$ rely only on the harmonic approximation of the energy, and can be obtained by minimizing the cost function
\begin{equation}
    \mathcal{C}(\vec{z}) = \frac{(\mathcal{M}: \vec{z} \vec{z})^2 }{\sum_{\langle i,j \rangle} (\vec{z}_{ij} \cdot \vec{z}_{ij})^2 }
    \label{eqn:phm_cost_fcn}
\end{equation}
with respect to the field $\vec{z}$. Here the notation $\mathcal{M}: \vec{z} \vec{z}$ denotes a double-contraction of the field $\vec{z}$ with the Hessian, sum in the denominator is over all interacting particle pairs $ij$, and $\vec{z}_{ij}\! =\! \vec{z}_j \!-\! \vec{z}_i$. Fields $\vec{\pi}$ associated with low-lying minima of $\mathcal{C}(\vec{z})$ satisfy $\partial {\cal C}/\partial\vec{z}|_{\vec{\pi}}\!=\!\zerovector$; they feature small stiffness (given by the numerator of Eq.~\ref{eqn:phm_cost_fcn}), and small participation ratio (thereby maximizing the denominator) \cite{richard_simple_2021}. Thus, long-wavelength plane waves are suppressed and disorder-related excitations (QLMs) are clearly identified. In practice, PHMs can be obtained by starting with an initial guess $\vec{z}_0$ and minimizing the cost function via a routine such as conjugate gradient. This highlights a major benefit of the bound operator approach - namely, it does not require a guess for the initial condition for each mode and thus recovers the full spectrum. For further information regarding PHMs, see references \cite{richard_simple_2021,kapteijns_nonlinear_2020}.

Now, if we use a particle displacement mode as the initialization for computing a PHM, $\vec{z}_0\!\rightarrow\!\vert D \rangle$, we obtain $\vert D_{PH} \rangle$, a $Nd$-dimensional 
\textit{pseudo-harmonic displacement mode} that corresponds to $\vert D \rangle$. In other words, $\vert D_{PH} \rangle$ should be viewed as $\vert D \rangle$ with obscuring de-localized features removed by the PHM protocol. 

\paragraph{Pseudoharmonic force modes} The forces that result from displacing the system via $\vert D_{PH} \rangle$ are simply given by $\vert F_{PH} \rangle = \mathcal{M} \vert D_{PH} \rangle$, which we thus define as \textit{pseudo-harmonic force modes}. We reemphasize that in the analysis presented here, we will use comparisons between $\vert D \rangle$ and $\vert D_{PH} \rangle$ (and $\vert F \rangle$ and $\vert F_{PH} \rangle$) to determine how close particle displacement and particle force modes are to their pure QLM-counterparts. 

\paragraph{Spatial decay} In general, it has been noted in Refs \onlinecite{gartner_nonlinear_2016, lerner_statistics_2016, lerner_micromechanics_2016} that the spatial structure of QLMs is characterized by a disorderd (dipole-response-like) core, decorated by a decaying field that scales as $\sim r^{-(\!\!\dbar-1)}$ where $r$ is the distance from the core. Since $\vert D_{PH}\rangle$ is a faithful representation of a QLM, we expect that its decay will match this profile. We can similarly examine the structure of $\vert F_{PH} \rangle$. Using arguments related to the definition of pseudoharmonic modes as fields minimizing Eq.~\ref{eqn:phm_cost_fcn} (see Refs.~\onlinecite{richard_simple_2021, lerner_micromechanics_2016}), we predict that pseudoharmonic force modes decay spatially as the cube of the gradient of $\vert D_{PH} \rangle$. Thus, $\vert \: \vert F_{PH} \rangle \vert \sim r^{-9}$ for $\dbar = 3$.

\subsubsection{Results: spatial structure and correspondence to PHMs}
As discussed above, we can use the formalism of the bond operator to study a family of disorder-related modes that represent structural precursors to plasticity in model glasses. Here, we discuss the resemblance of coordinate space representations of bond operator eigenvectors to their corresponding PHMs. For ease of visualization, Fig.s \ref{fig:2D_modes_decay_2}a-c show high-$\lambda$ modes derived from the bond operator in a representative 2D IPL system with $N\!=\!4096$ particles. The bond force mode $\vert f \rangle$ depicted in Fig.~\ref{fig:2D_modes_decay_2}a shows a field of extensions/compressions localized on a core of bonds. We note here that bond force modes can be interpreted as local strain fields due to the bond operator's relationship to the elastic Green's function as noted above. This set of bond forces translates to a particle force mode $\vert F \rangle$ and pseudo-harmonic force mode $\vert F_{PH} \rangle$ that are similarly localized, as shown in Fig.~\ref{fig:2D_modes_decay_2}b. Fig.~\ref{fig:2D_modes_decay_2}c shows the particle displacement mode $\vert D \rangle$ and pseudo-harmonic displacement mode $\vert D_{PH} \rangle$ that constitute the (linear) response of the system to the applied forces $\vert F \rangle$ and $\vert F_{PH} \rangle$ respectively. Significantly, we see that $\vert D \rangle$ closely resembles $\vert D_{PH} \rangle$. These displacement modes illustrate typical features of QLMs: low-stiffness excitations that localize on a distinct group of particles and have spatial structures characterized by a disordered core and radially decaying field.

We will now quantify the spatial structure and the convergence of $\vert F \rangle$ and $\vert D \rangle$ to their pseudo-harmonic counterparts. Fig.~\ref{fig:2D_modes_decay_2}d shows the decay profiles of the coordinate space modes discussed above. We computed the running median of the magnitude of polarization vectors on particles in each mode as a function of radius $r$ away from the core of the mode, where the core position is approximated by the location of the particle with the largest polarization vector magnitude. The results were averaged over 1000 high-$\lambda$ modes from 3D, IPL systems with $N\! =\! 2000$ particles, and $T_p\! =\! 0.45$.  As is expected for QLMs, $\vert D \rangle$ and $\vert D_{PH} \rangle$ decay spatially as $\sim\! r^{-(\!\!\dbar-1)}\!=\!r^{-2}$ in 3D~\cite{gartner_nonlinear_2016}. More surprisingly, $\vert F \rangle$ falls off with the same scaling, indicating that the forces prescribed by the high-$\lambda$ bond operator mode behave similarly to their corresponding displacements. This behavior of $\vert D \rangle$ and $\vert F \rangle$ is reminiscent of low-frequency harmonic modes, where the force and displacement modes are in the same direction. Conversely, $\vert F_{PH} \rangle$ decays much faster than $\vert F \rangle$: $\vert \: \vert F_{PH} \rangle \vert \!\sim\! r^{-9}$, in agreement with our prediction. This difference in spatial decay is consistent with the clear variation in general structure between the two types of particle force modes. Significantly, we have shown that high-$\lambda$ modes of the bond operator are robust representations of phonon-free QLMs in model glasses, highlighting $\mathcal{A}$ as a reliable source of useful micromechanical information.

Fig.~\ref{fig:2D_modes_decay_2}e shows the convergence of $\vert D \rangle$ to the structure of $\vert D_{PH} \rangle$ and $\vert F \rangle$ to $\vert F_{PH} \rangle$ respectively, as a function of $\lambda$. We compute and average the overlap of these pairs of modes (simply defined as $1\! -\! \langle V \vert V_{PH} \rangle$ for $V\! \rightarrow\! D, F$) for the same data as in Fig.~\ref{fig:2D_modes_decay_2}d. In the limit of large $\lambda$, $\vert D \rangle$ converges quickly as $\sim\! \lambda^{-4}$ to its PHM, whereas $\vert F \rangle$ does not. This suggests that small variations between $\vert D \rangle$ and $\vert D_{PH}\rangle$ give rise to large differences in the according force modes. In other words, there is some allowable variation in sets of applied forces that could still give rise to QLM-like displacement fields.

\subsection{Effect of thermal annealing on QLM properties reflected by $\mathcal{A}$}
The results presented in Fig.~\ref{fig:operator_spectrum_1} showed that the statistics of QLMs in glassy samples can be cleanly observed in $\mathcal{A}$ spectra for a variety of $T_{\rm p}$. We emphasize that these features are generically difficult to extract from the density of modes of the Hessian due to system size and hybridization effects. As discussed thoroughly in Ref.~\onlinecite{rainone_pinching_2020}, there are three notable QLM features that vary drastically with deeper annealing (decreasing $T_{\rm p}$): depletion of the density $\mathcal{N}$ of disorder modes in the system; shrinking of the QLM lengthscale $\xi$; and stiffening of the potential energy landscape (typically measured by the frequency $\omega_{\rm g}$ of soft modes). Our results from analysis of the bond operator robustly support these claims. By fitting our distributions of $\mathcal{A}$ eigenvlaues from Fig.~\ref{fig:operator_spectrum_1}a to the scaling relation $P(\lambda)\! \sim\! A_{\rm g} \; \lambda^{-7/2}$ in the large eigenvalue regime associated with QLMs, we obtain the values of the prefactors $A_{\rm g}$ which vary with $T_{\rm p}$. As is discussed extensively in Ref.~\onlinecite{rainone_pinching_2020}, an integral over the density of modes (or bond operator eigenvalues) shows that $A_{\rm g}$ encodes important information about both the density $\mathcal{N}$ and stiffness $\omega_{\rm g}$ of glassy QLMs. 

\begin{figure}[ht]
\includegraphics[width = 1.0\columnwidth]{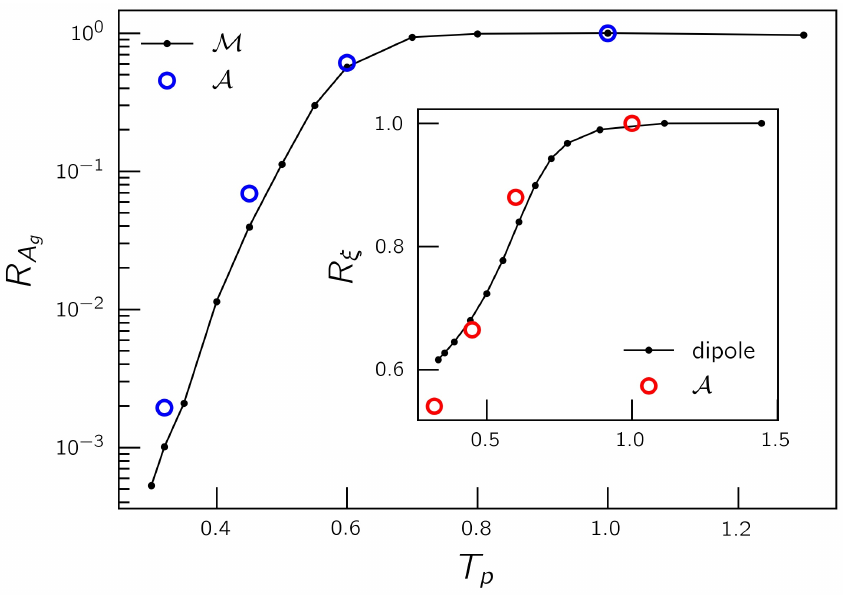}
\caption{Effect of thermal annealing on $A_{\rm g}$ and $\xi$. Prefactor ratios $A_{\rm g}$ extracted from $P(\lambda)$ distributions in Fig.~\ref{fig:operator_spectrum_1}a as a function of parent temperature (open blue circles), overlaid on data reproduced from Ref.~\onlinecite{rainone_pinching_2020} (black connected dots). The inset shows the QLM length scale ratio $R_\xi$ extracted from the participation plateaus in Fig.~\ref{fig:operator_spectrum_1}b as a function of parent temperature (open red circles), also overlaid on data from Ref.~\onlinecite{rainone_pinching_2020} (black connected dots).}
\label{fig:annealing_3}
\end{figure}

In their recent work, Rainone et.~al.~used a characteristic QLM frequency $\omega_{\rm g}$ extracted from the average response of glassy samples to applied dipolar forces to disentangle contributions to the $D(\omega)\! \sim \!\omega^4$ prefactor by $\mathcal{N}$ and a multiplicitive factor of $\omega_{\rm g}^{-5}$ that arises from integrating the density of modes over a finite range \cite{rainone_pinching_2020}. The authors then study the effect of annealing on QLM depletion by measuring $\mathcal{N}$ as a function of $T_{\rm p}$. Fig.~\ref{fig:annealing_3} shows our extracted $P(\lambda)$ prefactors as a function of parent temperature overlaid on the data reproduced from \cite{rainone_pinching_2020, lerner_characteristic_2018}. To remove the overall scale that differs between the two datasets, we divide the $P(\lambda)$ prefactors by that of the high $T_{\rm p}$ ensemble, and report values of the ratio $R_{A_{\rm g}}\! =\! A_{\rm g}/A_{\rm g, T_p \rightarrow \infty}$. As we see, there is good agreement between the observed trend in both studies. Thus, the bond operator spectrum is a reliable way to recover $A_{\rm g}$ as a function of annealing. Notably, the values of $A_{\rm g}$ vary by three orders of magnitude within the range of $T_{\rm p}$ that we explored. 

Next, we examine the effect of annealing on the QLM length scales extracted from the participation plateaus in Fig.~\ref{fig:operator_spectrum_1}b via Eq.~\ref{eqn:qlm_lengthscale}. We compare these lengths, shown in the inset of Fig.~\ref{fig:annealing_3}, to those reported in Ref.~\onlinecite{rainone_pinching_2020}. Rainone et.~al.~computed their QLM length scales for each sample parent temperature via the relation $\xi \!=\! 2 \pi c_s/\omega_{\rm g}$ where $\omega_{\rm g}$ is the dipolar response-derived characteristic frequency and $c_s$ is the shear wave speed. Again, we divide the $\xi$ values by the high temperature case and report the ratio $R_{\xi}\!\equiv\! \xi/\xi_{T_p \rightarrow \infty}$ for ease of comparison. Interestingly, despite seemingly different approaches for identifying the QLM length scales, our results agree quite well. These results further elucidate the utility of the bond operator to provide important insight into the properties of glassy QLMs a function of material preparation/memory.

\section{\label{sec:discussion}Summary, discussion and outlook}

Overall, we have shown that the bond force response operator, $\mathcal{A}$, is a powerful object that cleanly provides insight into many statistical and structural features of glassy instabilities. In particular, the spectrum of $\mathcal{A}$ is system-size independent, and appears to efficiently de-hybridize low-frequency QLMs from phononic excitations. We explored the spectral properties of the bond operator and highlighted the direct correspondence between its eigenmodes and QLMs. By examining ensembles of IPL model glasses we further explored the effect of deep annealing on QLM properties. Reliably identifying and understanding glassy defects is an important step toward building a comprehensive theory of the rheology of disordered materials. 

In this work we focused solely on annealing as a preparation protocol that effects the mechanical features of the resulting material. However, given the bond operator's strength as a tool for studying QLMs and their properties, it would be illuminating to utilize it future analyses of a broad range of computer glass ensembles. In particular, thoroughly studying the (phonon-free) vibrational spectra of loosely compressed particle packings near the unjamming transition would help inform current models for glassy instabilities.

We note that computing and diagonalizing the bond operator for large systems remains challenging, as fully diagonalizing the Hessian is computationally intensive. However, we provide a useful framework for analyzing the properties of disordered solids for small systems, for example, packings of a few thousand particles whose typical interaction range does not exceed a few particle diameters. Still, it may be possible to extract the same information that we have presented here using an approximation of the bond operator such as from partial sum over the low-frequency modes of the Hessian as in Eq.~\ref{eqn:A_hess_sum}; this is an avenue for future work. Furthermore, the bond operator could provide valuable insight into the structure of a variety of glass-forming models such as Stillinger-Weber or sticky spheres \cite{richard_simple_2021,Karina_sticky_spheres_part1_pre_2021}.

In this work we explored the spectrum of $\mathcal{A}$, which acts on bond space vectors and gives the deformation induced in particle bonds in response to a set of applied compressions/extensions. We provide intuition and numerical evidence for why a bond-space perspective on instabilities in disordered solids is valuable. In contrast to the spectrum of vibrational modes of the Hessian which is complicated by finite size effects and hybridizations with phononic excitations, the bond operator cleanly captures the statistics of disorder-related modes in model glasses. The high-eigenvalue scaling of the bond operator that we have measured, $P(\lambda)\! \sim\! \lambda^{-7/2}$, is equivalent to the universal $D(\omega)\! \sim\! \omega^4$ law \cite{kapteijns_universal_2018, LB_modes_2019, rainone_pinching_2020, modes_prl_2020}. 

As we have discussed, existing frameworks such as NPMs and PHMs have successfully identified populations of QLMs in disordered solids \cite{kapteijns_nonlinear_2020,richard_simple_2021}. Since NPMs require the computation of high order derivatives of the potential energy of the system, and both methods require initial guesses to find representations of QLMs via minimization of a nonlinear cost function, a relative strength of $\mathcal{A}$ is that it provides direct access to the full spectrum of phonon-free QLMs. We have shown that particle displacement modes $\vert D \rangle$, derived directly from high-$\lambda$ modes of the bond operator, quickly converge structurally to their corresponding PHMs.

Previous work used the response of disordered packings to local force perturbations (applied dipolar extension) as well as the density of harmonic modes $D(\omega)$ to determine the annealing dependence of QLM depletion, shrinking, and stiffening \cite{rainone_pinching_2020,rainone_statistical_2020,lerner_characteristic_2018}. In our analysis of the bond operator, we measured $P(\lambda)$ prefactors $A$ as a function of parent temperature $T_{\rm p}$ and found good agreement with the results of Ref.~\onlinecite{rainone_pinching_2020}. Additionally, we used the high-$\lambda$ plateau observed in the participation profile of bond operator modes to identify a length scale $\xi$ that decreases with deeper annealing. Conveniently, by computing only the bond operator, we gain access to spatial, mechanical, and statistical properties of glassy excitations that are important to popular theoretical frameworks such as STZ theory, SGR, and elastoplastic models.


\begin{acknowledgments}
E.L. acknowledges support from the NWO (Vidi grant no. 680-47-554/3259). J.A.G. and M.L.M. acknowledge support from the Simons Foundation grant \#454947 and National Science Foundation NSF-DMR-1951921. D.~R.~acknowledges support of the Simons Foundation for the ``Cracking the Glass Problem Collaboration" Award No.~348126.
\end{acknowledgments}

\appendix

\section{\label{sec:append_modes} Intermediate-$\lambda$ modes of $\mathcal{A}$ } 

\begin{figure}[htb]
\includegraphics[width = 1.0\columnwidth]{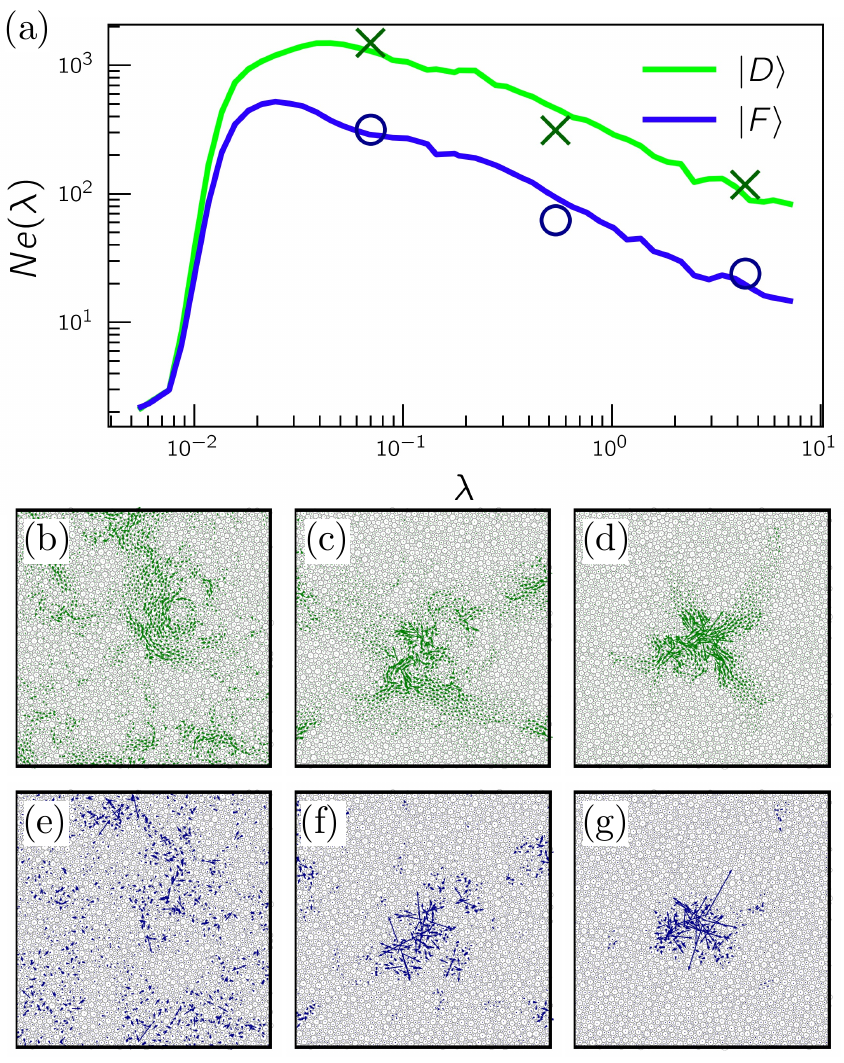}
\caption{\textbf{a)} Participation ratio as a function of $\mathcal{A}$ eigenvalue $\lambda$ for particle force modes (blue, bottom) and particle displacement modes (green, top). Markers represent the participations and eigenvalues associated with the sample $\vert D \rangle$ and $\vert F \rangle$ modes in (b) - (g), which are arranged in the same order. \textbf{b)} - \textbf{d)} particle displacement modes corresponding to the profile in (a), arranged from low to high $\lambda$ from left to right. \textbf{e)} - \textbf{g)} particle force modes corresponding to the profile in (a), arranged from low to high $\lambda$ from left to right.}
\label{fig:append_example_modes}
\end{figure}

Similarly to Fig.~\ref{fig:operator_spectrum_1}b in the maintext, in Fig.~\ref{fig:append_example_modes}a we show the participation ratio (multiplied by system size) as a function of $\lambda$ for a small ensemble of particle force modes and particle displacement modes derived from eigenvectors of $\mathcal{A}$. The data presented here is from $\sim\!150$ IPL glasses in 2D with $N\! =\! 4096$ particles and $T_{\rm p} \!=\! 0.7$. We see that the participation profiles for the two types of modes vary by an overall factor for most of the range in $\lambda$, but the trends and tendency toward a plateau are the same. In panels (b)-(d) and (e)-(g) of Fig.~\ref{fig:append_example_modes}, we show example $\vert D \rangle$ and $\vert F \rangle$ modes respectively that lie in the intermediate-$\lambda$ regime of the spectrum. Approaching higher $\lambda$, the modes begin to resemble QLMs. 

\section{\label{sec:append_system_size} $\mathcal{A}$ spectrum and system size dependence}

\begin{figure}[htb]
\includegraphics[width = 1.0\columnwidth]{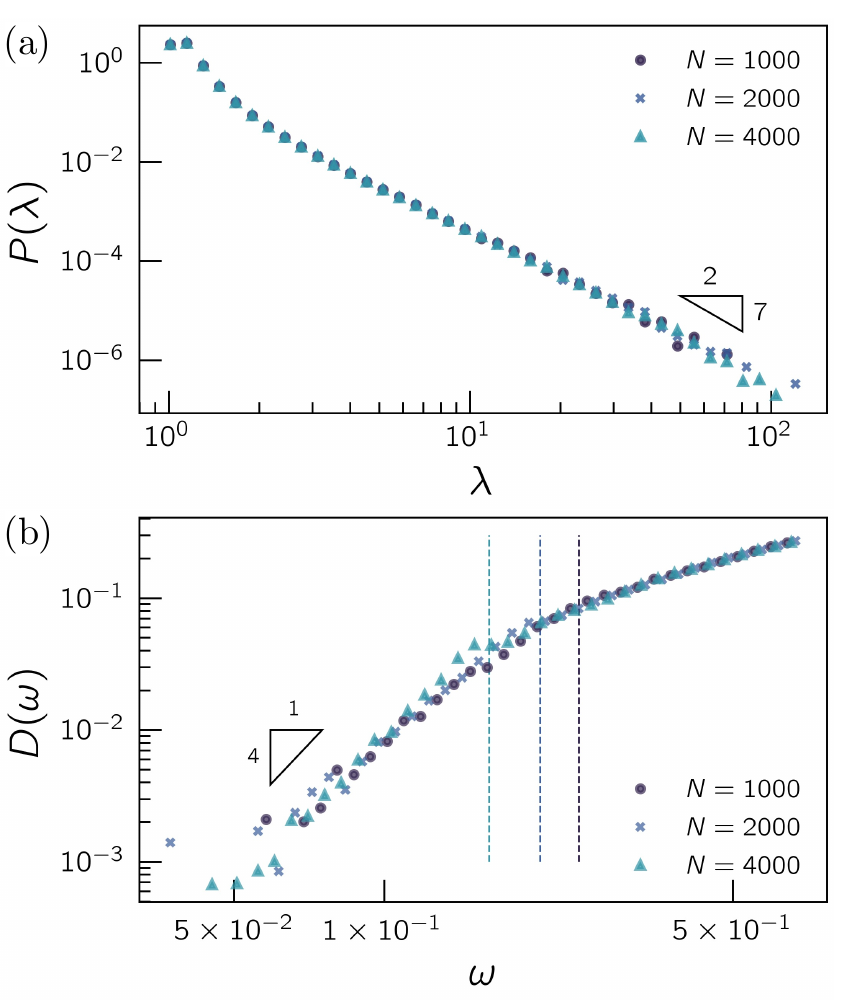}
\caption{$P(\lambda)$ and $D(\omega)$ distributions for ensemble of harmonic sphere packings. Spectra for varying system sizes are represented with different shape/color of marker as shown. \textbf{a)} Full $P(\lambda)$ spectrum, where the scale triangle represents $P(\lambda)\! \sim\! \lambda^{-7/2}$ scaling. \textbf{b)} The low-frequency $D(\omega)$ is shown to emphasize the presence of phonons. The scale triangle represents $D(\omega)\! \sim\! \omega^4$ scaling. Dashed vertical lines depict the lowest phonon frequency given by $\omega_{\text{ph}}\! =\!2 \pi c_s/L$ where corresponding peaks are visible.}
\label{fig:append_N_spectra}
\end{figure}

To show that $\mathcal{A}$'s eigenspectrum is independent of system size, we analyzed a variety of 3D 50:50 bidisperse harmonic sphere packings prepared at pressure $p\! =\! 0.1$ expressed in simulational units. The ratio of diameters of large and small spheres is set to 1.4. The harmonic spheres interact via the pairwise potential
\begin{equation}
    u(r_{ij}) =
    \begin{cases}
        \frac{1}{2}k(r_{ij}-l_{0,ij})^2, \: \: r_{ij} \leq l_{0,ij}\\
        0, \: \: r_{ij} > l_{0,ij}
    \end{cases}
    \label{eqn:pair_energy_harm}
\end{equation}
where $k$ (set to unity) is a spring constant associated with the bond (particle pair) $ij$ and $l_{0,ij}$ is the bond rest length. All $l_{0,ij}$ are set to the sum of the radii of particles $i$ and $j$, $l_{0,ij}\! =\! d_i\! + \!d_j$. We created glasses at the desired target pressure using the FIRE energy minimization algorithm \cite{bitzek_structural_2006}. See Ref.~\onlinecite{lerner_mechanical_2019} (discussion of Hertizan sphere packing preparation) for further details of the implementation. Here, we analyze 1000 independent configurations for each $N\! \in\! \left\{1000, 2000, 4000 \right\}$. 

By comparing panels (a) and (b) of Fig.~\ref{fig:append_N_spectra}, we see that $P(\lambda)$ is independent of system size, while $D(\omega)$ has phonon peaks in the spectrum that change frequency as a function of simulation box size $L$ (such a peak is also clearly visible in Fig.~\ref{fig:example_hessian_bond} in the main text). This emphasizes the utility of the bond operator for investigating the phonon-free spectrum of QLMs in model glasses. 

\vspace{1cm}

\bibliography{Bond-space_operator}

\end{document}